\documentclass{article}
\usepackage[utf8]{inputenc}
\usepackage{amsmath,amssymb}
\usepackage[margin=1in]{geometry}
\usepackage{graphicx}
\usepackage{color}
\usepackage{ulem}
\usepackage{marginnote}
\usepackage{url}
\usepackage{authblk}

\DeclareUnicodeCharacter{0301}{\'{e}}
\DeclareUnicodeCharacter{03B1}{$\alpha$}

\title{Escape by jumps and diffusion by $\alpha$-stable noise across the barrier in a double well potential}

\author[1]{Ignacio del Amo}
\author[2]{Peter Ditlevsen}

\affil[1]{University of Exeter, Exeter, Devon EX4 4PY, United Kingdom}
\affil[2]{Niels Bohr Institute, University of Copenhagen, Copenhagen, DK-1165, Denmark}

\date{December 2022}

\begin{document}

\maketitle

\Large
\begin{abstract}
Many physical and chemical phenomena are governed by stochastic escape across potential barriers. The escape time depends on the structure of the noise and the shape of the potential barrier. By applying $\alpha$-stable noise from the $\alpha=2$ Gaussian noise limit to the $\alpha<2$ jump processes, we find a continuous transition of the mean escape time from the usual dependence on the height of the barrier for Gaussian noise to a dependence solely on the width of the barrier for $\alpha$-stable noise. We consider the exit problem of a process driven by $\alpha$-stable noise in a double well potential. We study individually the influences of the width and the height of the potential barrier in the escape time and we show through scalings that the asymptotic laws are described by a universal curve independent of both parameters. When the dependence in the stability parameter is considered, we see that there are two different diffusive regimes in which diffusion is described either by Kramer's time or by the corresponding asymptotic law for $\alpha$-stable noise. We determine the regions of the noise parameter space in which each regime prevails, and exploit this result to construct an anomalous example in which a double well potential exhibit a different diffusion regime in each well for a wide range of parameters.
\end{abstract}

\section{INTRODUCTION}
Noise induced transitions between different stable states occur in many types of simple or complex systems: chemical reactions \cite{falk2017minimal}, geophysics and climate
modelling \cite{castellana2020noise,debussche:2013}, gene regulatory networks \cite{bosaeus2012tension, kirunda2021effects}, just to mention a few.  

Adding unbounded noise to a deterministic multistable system causes the stable states of that system to become metastable, keeping the process within the basin of attraction during long times, but not preventing entirely its escape to another nearby basin. 
Assuming the noise to be Gaussian is natural as the central limit theorem dictates that this is indeed the limiting distribution for summing infinitesimal increments from independent identically distributed (i.i.d.) stochastic variables with a finite variance. However, for heavy tail distributions, $\phi(x)$, where moments $\langle |x|^\beta\rangle =\int |x|^\beta\phi(x)dx$ are only finite for $0<\beta\le\alpha<2$, an extension of the central limit theorem states that summing i.i.d. increments from $\phi(x)$ results in universal distributions depending only on $\alpha$ and a skewness parameter, which we shall not be concerned with here.
Gaussian noise is thus the limit case when $\alpha=2$ in the wider category of stable noise distributions called $\alpha$-stable noises, where $\alpha$ is called the stability parameter and can take any value in $(0,2]$. Note that in the limit $\alpha=2$ the tail of the distribution is exponential and all moments are finite. Stability of a distribution is defined with respect to addition; the sum of two variables has the same distribution with suitable rescaling. For a brief account of the symmetric $\alpha$-stable processes see the Appendix.

\begin{figure}[htbp]
\begin{center}
\includegraphics[width=\textwidth]{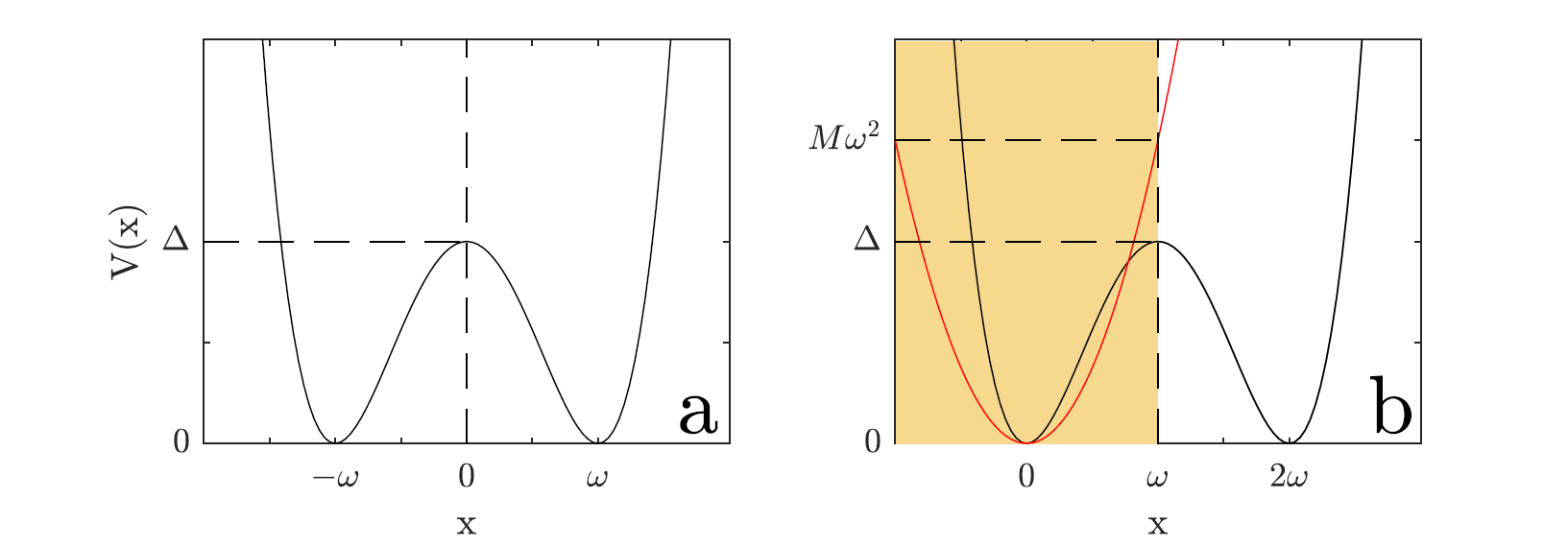}
\end{center}
\caption{\label{fig:potential} a. Illustration of the potential wells used and the well parameters. b. Analytic results for first exit time of the interval $(-\infty, \omega)$ (yellow area) in a single well harmonic potential is used as an approximation of the escape time across the barrier in the double-well potential $V(x)$. The first exit time does not depend on the steepness of the harmonic potential.}
\end{figure}

Here we consider a stochastic process described by a Langevin equation:
\begin{align}\label{eq:Langevin}
    dX=-\frac{dV(x)}{dx}dt + \frac{\sigma}{\sqrt{2}} dL_\alpha
\end{align}
which is a simple one dimensional gradient system, $\sigma$ is the intensity of the noise, and $dL_\alpha$ is the differential of an $\alpha$-stable process.
$V(x)$ is a double well potential,
\begin{align} \label{eq:Potential}
    V(x)= \Delta (1-x^2/\omega^2)^2,
\end{align}
which is a quartic polynomial with two wells, placed in $x=-\omega$ and $x=\omega$ and a local maximum between them on $x=0$ of height $\Delta$. 
(Fig.~\ref{fig:potential}a).

We shall use the slightly unconventional notation $dL_2=dB$ in order to also cover the Gaussian case (Brownian motion). 
The constant $1/\sqrt{2}$ in Eq.~\eqref{eq:Langevin} is there to normalize the noise in the case $\alpha=2$ to a standard normal distribution, given our choice to simulate the noise using the direct method for symmetric $\alpha$-stable noise described in \cite{janicki2021simulation}.

Evidence of naturally generated $\alpha$-stable processes exists in the paleoclimatic record \cite{ditlevsen:1999}, and as such it has applications in geophysics and climate modelling \cite{lucarini:2022,debussche:2013}. However the interest in $\alpha$-stable processes spans more fields, with applications as well in simulation of financial markets \cite{delong2006optimal}, animal behaviour \cite{viswanathan1996levy}, epidemiology models \cite{cui2022impact,ding2021stochastic} or stochastic homogenization of intermittent systems \cite{gottwald:2021}, to cite some.

The Gaussian process is continuous and here referred to as a diffusion process, while 
 $\alpha$-stable processes, with $\alpha <2$, are continuous in probability, but not continuous. That implies that there is a zero measure set of discontinuities (jumps) in the path of the process. The discontinuities dominate the process when the intensity of the noise is low. These processes with a mixture of diffusion and jumping are also referred to as L\'{e}vy flights.

When escaping across the potential barrier separating the two potential wells, the drift from the gradient of the potential is working against the diffusion, thus the escape time depends on the height of the barrier (Kramer's law). This is in contrast to a jump process where the jumps in the trajectory are not affected by the potential barrier that separates stable states. In this case we should rather expect the escape time to depend on the width of the barrier to be overcame by single jumps. Because of this, in the weak noise limit the transition times between stable states are strongly dependent on the height of the barrier when $\alpha=2$, but independent from it when $0<\alpha<2$. The goal of this study is to describe the transition from a height dominated escape time when the noise is Gaussian, to a width dominated one when the process is driven by $\alpha$-stable noise.

\section{SIMULATING THE PROCESS}
When it comes to simulating the $\alpha$-stable process in Eq.~\eqref{eq:Langevin}, there is a slight inconsistency between the analytical results for escape times, the numerical algorithms to solve this equations and the systems which are applied to. The results for the escape time rates of Eq.~\eqref{eq:Langevin} require a potential well with only one stable state and that grows faster than linearly on infinity \cite{imkeller:2006,imkeller2010hierarchy}. However, most numerical algorithms require, for convergence, at least a one-sided Lipschitz condition on the drift term (i.e. that the drift grows at most linearly at $\pm\infty$) \cite{dareiotis:2016,kelly:2018,fang:2020}. Moreover, the Euler-Maruyama method for SDEs with a cubic drift is proven to not converge due to the non-boundedness of the finite time increments \cite{gottwald:2021}. For a quartic double well potential, neither of these requirements apply, since it has multiple stable states and does not meet the one sided Lipschitz condition.  For these reasons, we employ a tamed Euler method \cite{gottwald:2021,kelly:2018}, and we check our results against the globally Lipschitz drift given by 

\begin{align}\label{eq:LipschitzPotential}
    V_{Lip}(x)=\Delta\frac{\sqrt[4]{N^4+(1-x^2/\omega^2)^2}-N}{\sqrt[4]{N^4+1}-N},
\end{align}
where we use $N=200$. Note that this potential has the same local maximum and minima as Eq.~\eqref{eq:Potential}, and a linear behaviour when $|x|\uparrow \infty$. In a reasonably large neighbourhood of $x=0$ the Lipschitz potential is very close to the quartic one:
\begin{align*}
    |V(x)-V_{Lip}(x)|< 0.023 \Delta,
\end{align*}
for $|x|\leq10\omega$. This allows us to distinguish if the behaviour of Eq.~\eqref{eq:Langevin} is determined by the shape of the potential near the wells or also by the rate of growth of the potential.

\section{ESCAPE TIME}
When $\alpha = 2$ the relevant average escape time for this system under weak noise is given by Kramer's law \cite{berglund2006noise}: 
\begin{align}\label{eq:Kramers}
    \lambda_2(\sigma)= \frac{\pi}{\sqrt{V''(y^*)|V''(x^*)|}} e^{\frac{2(V(x^*)-V(y^*))}{\sigma^2}}
\end{align}
where $\lambda$ is the mean escape time, $x^*$ is the maximum of the potential barrier and $y^*$ the stable fixed point whose basin contains the process before the escape. With the potential given by Eq.~\eqref{eq:Potential}, Kramer's time is
\begin{align}\label{Eq:Kramerspotential}
    \lambda_2(\sigma)=\frac{\pi \omega^2}{\Delta\sqrt{8}} e^{\frac{2\Delta}{\sigma^2}}.
\end{align}

A good approximation to the mean escape time when $\alpha<2$ is obtained from \cite{imkeller:2006}. In that paper, an analytic expression for the first exit time of the interval $(-\infty, \omega)$ from a harmonic potential $V_{harm}(x)=M x^2$, with the single minimum at $x=0$ in the weak noise limit is derived, finding it equal to 
\begin{align*}
    \alpha w^\alpha \varepsilon^{-\alpha},
\end{align*}
with $w$ being the distance between the minimum and the edge of the interval we want to escape from, and $\varepsilon$ is the noise intensity.
To obtain the correct asymptotic law for our Langevin equation we follow the transformations outlined in \cite{Chechkin:2007}. The noise intensity $\varepsilon$ employed in \cite{imkeller:2006} is not equivalent to $\sigma$. First we define the normalized noise amplitude $\sigma'=\sigma/\sqrt{2}$.
We have that
\begin{align*}
    \sigma = \mathcal{C}(\alpha)^{1/\alpha} \varepsilon
\end{align*}
where 
\begin{align*}
    \mathcal{C}(\alpha)= \frac{2\Gamma(1-\alpha)}{\alpha} \cos{\frac{\pi\alpha}{2}}.
\end{align*}
This is due to different definition of the $\alpha$
-stable noise used in \cite{imkeller:2006}. 
From this we obtain 
\begin{align*}
    \alpha w^\alpha \varepsilon^{-\alpha} =  2^{(2-\alpha)/2} \Gamma(1-\alpha) \cos\left({\frac{\pi\alpha}{2}}\right)
    \omega^\alpha \sigma'^{-\alpha},
\end{align*}
by substituting $\varepsilon$,  $\mathcal{C}(\alpha)$, $\sigma$ and $w=\omega$, as the correct escape time law from $V_{harm}$, for a Langevin equation forced with normalized $\alpha$-stable noise. In the remaining we drop the $'$ from the $\sigma$, and we set the dimensionless constant $K_\alpha$ to 
\begin{align*}
    K_\alpha= 2^{(2-\alpha)/2} \Gamma(1-\alpha) \cos\left({\frac{\pi\alpha}{2}}\right),
\end{align*}
for convenience, leaving the formula for the escape time as
\begin{align}\label{eq:LevyEscapeTimePotential1}
    \lambda_\alpha(\sigma) = K_\alpha \omega^\alpha \sigma^{-\alpha}.
\end{align}
Note that the exit time does not depend on the steepness of the potential, $M$ (see Fig. \ref{fig:potential}b.).

By simulation we have verified Eq. \eqref{eq:LevyEscapeTimePotential1} as an expression also for the mean escape time across the potential barrier for the double-well potential. However, for numerical stability we set the threshold to escape a basin to be when the process surpasses $1.9\omega$ in the direction of the other well. This introduces a rescaling of the width, which we have found numerically that translates into a factor of $(1.95)^\alpha$ in the escape time, which we include in the dimensionless constant $K_\alpha' = (1.95)^\alpha K_\alpha$, and again, drop the $'$ to remain with Eq.\ref{eq:LevyEscapeTimePotential1} as the correct escape time.

\section{SCALING THE POTENTIAL} \label{sec:scaling}

Observing the scaling $dL_\alpha\sim dt^{1/\alpha}$, fully analogous to the Brownian motion case $dB\sim dt^{1/2}$, it is straightforward to verify that Eq.~\eqref{eq:Langevin} is invariant under the rescaling, $(\tilde{t}, \tilde{x}, \tilde{\sigma},\tilde{V}(\tilde{x}))=(\tau t, \xi x, \tau^{-1/\alpha}\xi\sigma, (\xi^2/\tau)V(x))$, thus we can directly obtain the mean escape time across the barrier for a rescaled potential, $\tilde{V}(\tilde{x})=sV(\xi x)$, by scaling time, $\tilde{t}=(\xi^2/s)t$:

\begin{align}
   \tilde{\lambda}_\alpha(\tilde{\sigma}, \tilde{V}(\tilde{x}))=(\xi^2/s)\lambda(\sigma, V(x))=(\xi^2/s)\lambda_\alpha(\xi^{(2-\alpha)/\alpha}s^{-1/\alpha}\tilde{\sigma}, V(x)).
\end{align}
Thus scaling the width of the barrier with a factor $\xi$, keeping the height of the barrier constant $(s=1)$ we get for $\alpha=2$:

\begin{equation}
\tilde{\lambda}_2(\sigma, V(\xi x))=\xi^2\lambda_2(\sigma, V(x)), 
\end{equation}
and for $0<\alpha<2$ the same relation holds:
\begin{equation}
\tilde{\lambda}_\alpha(\sigma, V(\xi x))=\xi^\alpha\lambda_\alpha(\xi^{(2-\alpha)/\alpha}\sigma, V(x))=\xi^\alpha \lambda_\alpha(\sigma, V(x)), 
\end{equation}
where Eq.~\eqref{eq:LevyEscapeTimePotential1} was used in the last equality.
Likewise, for a potential with fixed width $(\xi=1)$ and scaled height we get for $0<\alpha\le 2$:

\begin{equation}
\tilde{\lambda}_\alpha(\sigma,s V(x))=(1/s)\lambda_\alpha(\sigma/s^{1/\alpha},V(x)).
\end{equation}

For any $\alpha\in (0,2]$ scaling the height by $s$ has the effect of scaling  the mean escape time by $1/s$ and  the noise intensity by $1/s^{1/\alpha}$. Note that this is continuous in the limit when $\alpha\uparrow 2$. 

\begin{figure}[htbp]
\begin{center}
\includegraphics[width=12cm]{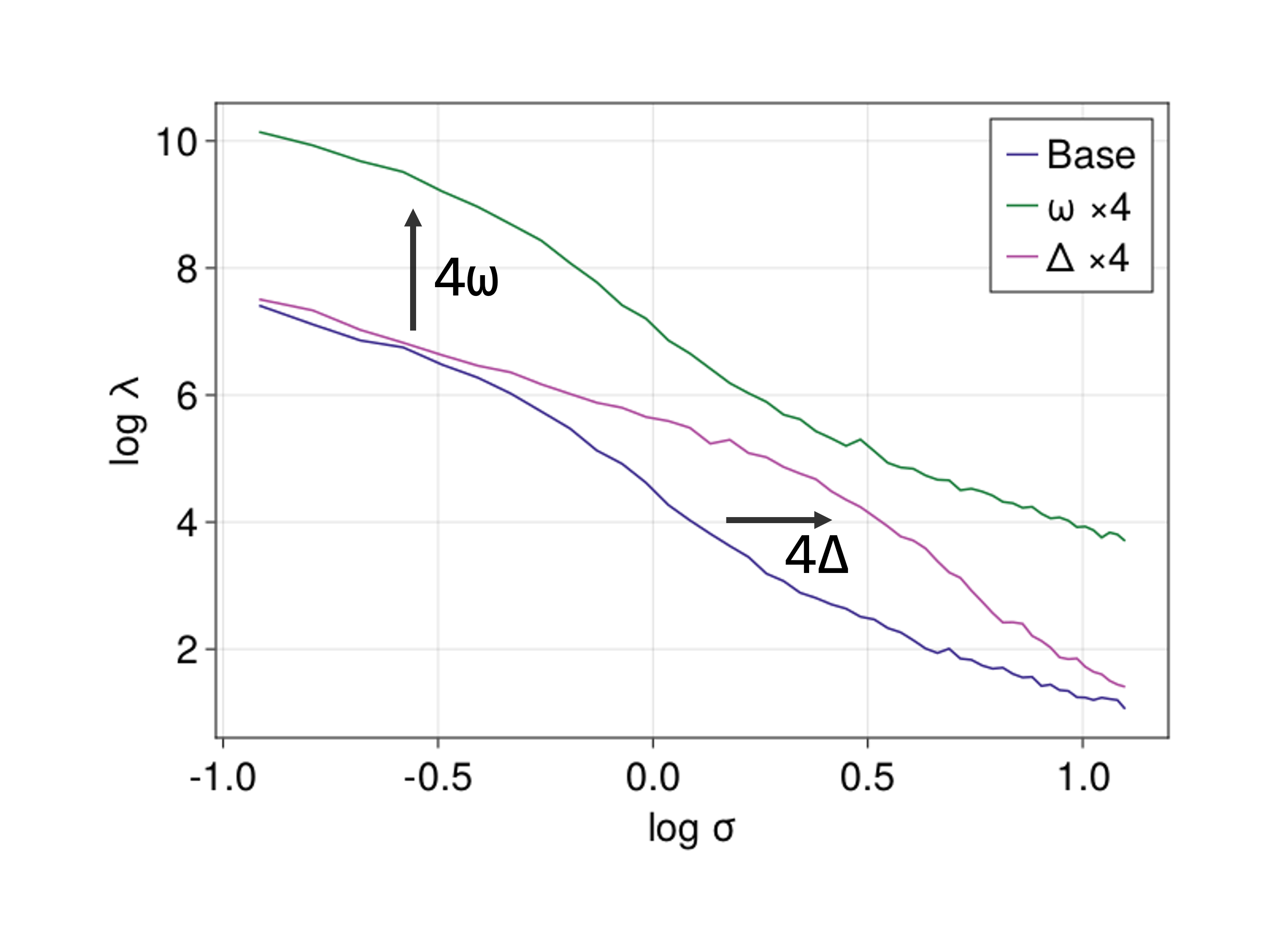}
\end{center}
\caption{\label{fig:scaling}This figure shows the effect of scaling the width $\omega$ or the height $\Delta$ in the average escape time. The mean escape time curves plotted correspond to a base case for a fixed value of $\omega$ and $\Delta$, a case corresponding to four times more width and fixed $\Delta$ and a case corresponding to four times more height and fixed $\omega$. Scaling the width ads a prefactor to the escape time, while scaling the height is equivalent to reescaling the noise intensity and the time in such a way that the asymptotic law remains invariant.}
\end{figure}

Fig.~\ref{fig:scaling} shows three mean escape time curves $\lambda_\alpha(\sigma)$ for the value $\alpha=1.95$ that show the effect of scaling the width $\omega$ and the height $\Delta$. 

\section{UNIVERSAL CURVE} \label{sec:universal}

\begin{figure}[htbp]
\begin{center}
\includegraphics[width=12cm]{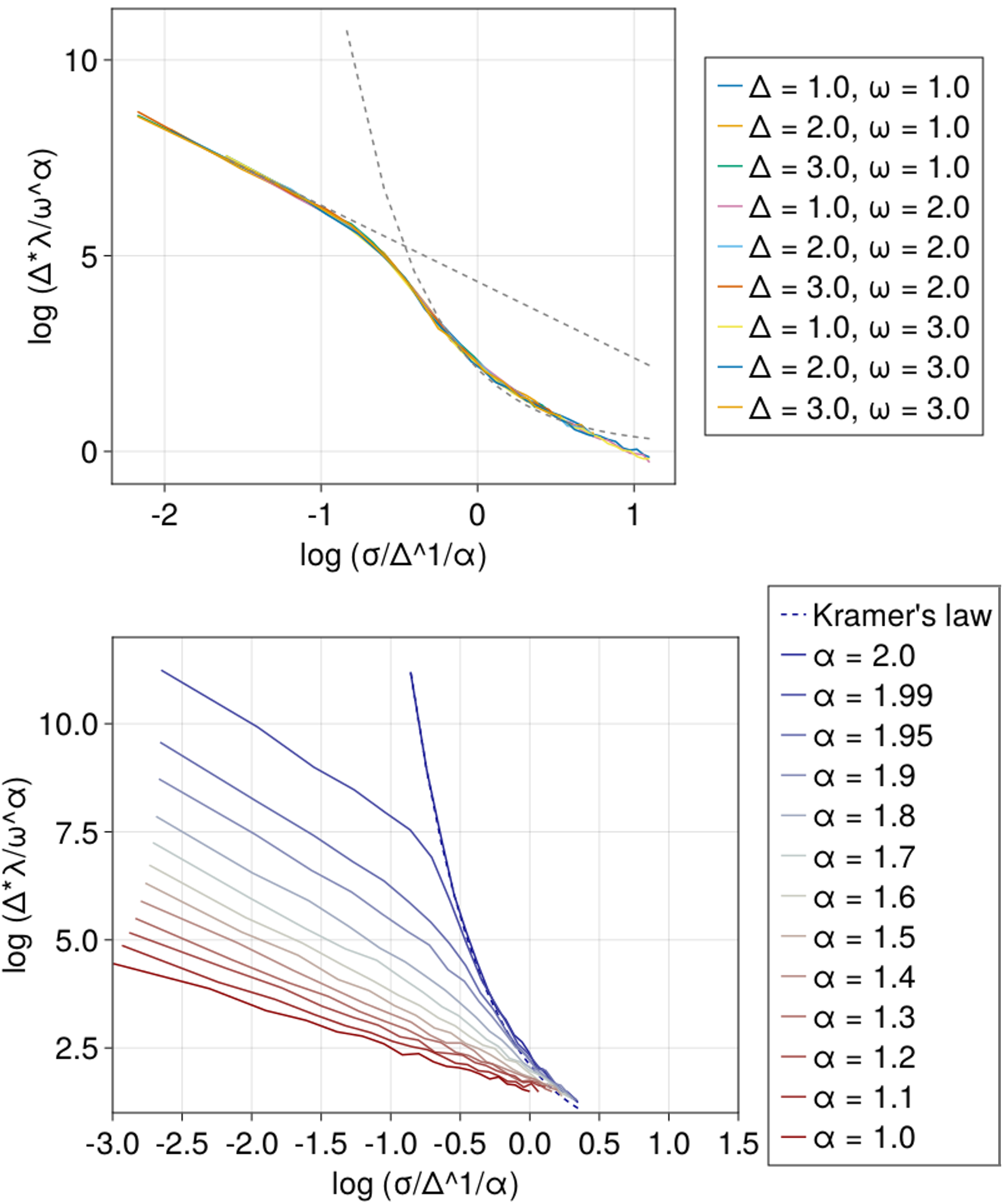}
\end{center}
\caption{\label{fig:universal} On the top, universal curve for $\alpha = 1.95$. The curves corresponding to different values of $\omega$ and $\Delta$ collapse into one curve in the axis scaled as in Eq.\ref{eq:universalAxes}. The dashed lines are the asymptotic escape times for the Gaussian case (Kramer's law) and the $\alpha$-stable case.
On the bottom, mean escape time curves for different values of the parameter $\alpha$, note that as $\alpha$ increases the curves .}
\end{figure}

In the last section we have seen how to express the mean escape time of a particle in a scaled potential well as a function of the scaling and the original escape time. This allows us to construct axes in which the escape time curves for different values of the width and height collapse into one single universal curve that describes the average waiting time for the whole family of potential wells parameterized by $\omega$ and $\Delta$. Any potential of this kind can be described by setting the width and the height of the well to 1 and choosing to scale the width by $\omega$ and the height by $\Delta$. Omitting for brevity $V$ as argument; $\lambda(\sigma)$ is the escape time curve for a potential well of half-width $\omega_0=1$ and height $\Delta_0=1$, then if $\tilde{\lambda}(\sigma)$ is the escape time curve for a potential well of width $\omega$ and height $\Delta$, we have: 
\begin{align}\label{eq:universalAxes}
   \tilde{\lambda}(\sigma)= (\omega^\alpha/\Delta) \lambda(\sigma/\Delta^{1/\alpha})
\end{align}
for any $\alpha\in (0,2]$. This is shown in Fig.~\ref{fig:universal}, where nine curves corresponding to potentials $V(\omega, \Delta)$ with  $\omega \in\{1, 2, 3\}$ and $\Delta  \in\{1, 2, 3\}$ are plotted together. The green curve corresponds to Kramers escape time \eqref{eq:Kramers} while the purple curve corresponds to the asymptotic escape time \eqref{eq:LevyEscapeTimePotential1}. Note that while we have parameterized the symmetric potential Eq.\ref{eq:Potential} using the symbols $\omega$ and $\Delta$, in reality this potential is just a $\omega$ stretching of the $x$ axis and a $\Delta$ stretching of the $y$ axis of the polynomial $(1-x^2)^2$. With this we want to emphasize that the results in the previous section and this one are general respect of the potential used, not particular of Eq.\ref{eq:Potential}.

\section{DIFFUSIVE REGIMES FOR THE SYMMETRIC POTENTIALS}\label{sec:diffusive}

With the scaling relation Eq.~\eqref{eq:universalAxes} describing the dependence on $\omega$ and $\Delta$, the only parameter determining the escape time as a function of $\sigma$ is the stability parameter $\alpha$. 

Fig.~\ref{fig:universal} shows the mean escape time curves that correspond to different values of $\alpha$. The darkest blue obtained for $\alpha=2$ matches Kramer's law (in dashed line), Eq.~\eqref{Eq:Kramerspotential}. In the large noise limit, this  dominates the escape times for all $\alpha$. This was already noted in \cite{chechkin:2005,Chechkin:2007}. 

For $\alpha$ close to, but smaller than, 2 we see a clear change of behaviour that corresponds to two different regimes. The curves follow a diffusive regime given by Eq.~\eqref{Eq:Kramerspotential} until the noise is weak enough to enter the weak noise asymptotic regime. Then the rate of crossing the barrier through one big jump becomes larger than the rate of crossing the barrier through an almost continuous path similar to a Gaussian crossing, and the system enters a jumping regime. 
Similarly, the escape time transitions from following a rapidly growing exponential law that depends mostly in the height of the barrier, to a slower growing power law that depends exclusively in the width of the barrier.

The transition from a height dominated law to a width dominated law takes place when the noise decreases to a given level that depends on the value of $\alpha<2$. This can be used to give a precise notion of ``weak noise'', as the amplitude of noise for which the system switches from one diffusive regime to the other. 
The Lévy–Khintchine theorem states that the $\alpha-$stable process is a compound process, which is a sum of a Brownian motion and a jump process. Thus for large noise intensity where the Brownian motion dominates, we may use Eq. \eqref{Eq:Kramerspotential} to estimate the mean waiting time, $\lambda(\sigma)$, while for small noise intensity, where the jump process dominates, we may use Eq. \eqref{eq:LevyEscapeTimePotential1} to estimate $\lambda(\sigma)$.
This weak noise threshold $\sigma_t(\alpha)$ is then simply the intersection point of the curves described by Eqs. \eqref{Eq:Kramerspotential} and \eqref{eq:LevyEscapeTimePotential1}. This is obtained as the solution to the implicit equation
\begin{align} \label{eq:weakNoise}
    K_{\alpha}\omega^\alpha\sigma_t(\alpha)^{-\alpha} =\frac{\pi\omega^2}{\Delta\sqrt{8}}e^\frac{2\Delta}{\sigma_t(\alpha)^2}.
\end{align}

This equation does not always have a solution for every $\alpha \in (0,2)$. We are only concerned with the behaviour of this equation in the first quadrant, particularly when the noise amplitude $\sigma$ is a small positive number and the stability parameter $0<\alpha<2$. In a log-log plot, the left-hand side is linear relationship, and the right-hand side is a power law. For any fixed $\alpha<2$, the right-hand side goes faster to infinity when $\sigma\downarrow 0$ than the left-hand side, and tends to a constant when $\sigma\uparrow\infty$, while the left-hand side tends to $-\infty$. This means that right hand side is above the left-hand side in the two limits of the first quadrant. When $\alpha\uparrow 2$, the term $\Gamma(1-\alpha)\cos(\pi\alpha/2)$ inside $K_\alpha$ tends to infinity, while the right-hand side is independent of $\alpha$, so for high enough $\alpha$, there will be at least two crossings. In some cases, the two rates can be tangent, giving a minimum value of $\alpha$ such that crossing occurs.

If it exists, the minimum $\alpha$ such that equation \ref{eq:weakNoise} has a solution can be found through the following method. Consider the inverse function $\alpha_t(\sigma)$ as the function such that $(\alpha_t(\sigma),\sigma)$ parameterizes the level curve defined by Eq.\ref{eq:weakNoise}.  Taking logarithms and the implicit derivative respect to $\sigma$ in this equation produces
\begin{align*}
    \frac{d}{d\sigma}\left(\log K_{\alpha_t(\sigma)} +\alpha_t(\sigma)\log(\omega/\sigma)\right)  = \frac{d}{d\sigma}\left(\log \frac{\pi\omega^2}{\Delta\sqrt{8}} +\frac{2}{\sigma^2}\right),
\end{align*}
since all considered quantities are positive. The equation can be solved for $\Dot{\alpha}_t(\sigma)$ yielding the differential equation

\begin{align} \label{eq:differentialAlpha}
    \Dot{\alpha}_t(\sigma)=\frac{\frac{\alpha_t(\sigma)}{\sigma}-\frac{4\Delta}{\sigma^3}}{\Tilde{K}_{\alpha_t(\sigma)}+\log (\omega/\sigma)}
\end{align}
where $\Tilde{K}_{\alpha_t(\sigma)}$ is the quantity such that $\frac{d}{d\sigma}\log K_{\alpha_t(\sigma)}=\Dot{\alpha}_t(\sigma)\Tilde{K}_{\alpha_t(\sigma)}$. Now assuming $\alpha_t(\sigma)<2$ and setting the derivative equal to zero we obtain the solution curve $\alpha_t^*(\sigma)=4\Delta/\sigma^2$. Together with Eq.\ref{eq:weakNoise} we can obtain an implicit equation for the minimum $\alpha^*$ such that there is an intersection for the two rates:
\begin{align}\label{eq:minalpha}
        K_{\alpha^*}\omega^{\alpha^*}\left(\sqrt{4\Delta/\alpha^*}\right)^{-\alpha^*} =\frac{\pi \omega^2}{\Delta\sqrt{8}} e^{\frac{\alpha^*}{2}}.
\end{align}
Since the terms that depend in the curvature of the potential disappear when taking the derivative, eq.~\ref{eq:differentialAlpha} should hold for potentials other than the family described by Eq.\ref{eq:Potential}. 

\begin{figure}
    \centering
    \includegraphics[width=12cm]{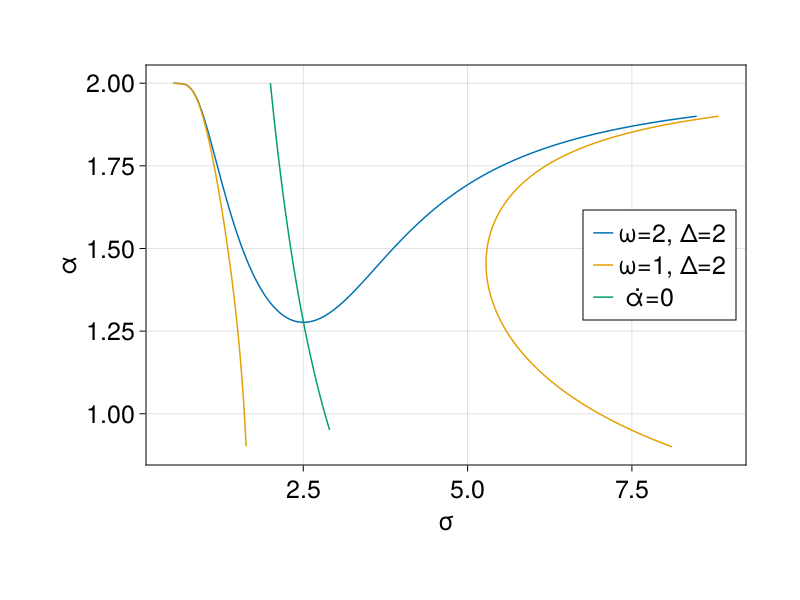}
    \caption{This plot shows two a numerical integration of the weak noise threshold (Eq.\ref{eq:weakNoise}) for each value of $\alpha$. The first case (blue) corresponds to the parameter values $\omega=2$ and $\Delta=2$. The minimum of $\alpha$ for this parameter values is $\alpha^*=1.27651564...$, where it intersects the curve $\Dot{\alpha} = 0$. The second (orange) corresponds to the parameter values $\omega=1$ and $\Delta=2$. For this parameter values there is no intersection and thus no minimum of alpha, so for any $\alpha\in (0,2)$ there are two solutions. }
    \label{fig:weaknoiseThreshold}
\end{figure}

Note that even though eq.~\ref{eq:differentialAlpha} is not defined when $\alpha_t(\sigma)=2$, as we take the limit $\alpha\uparrow 2$, $\Dot{\alpha}_t(\sigma)$ tends to zero. Figure~\ref{fig:weaknoiseThreshold} shows a numerical solution of the weak noise threshold $\sigma_t(\alpha)$ for the cases $\omega=2$, $\Delta=2$ and $\omega=1$, $\Delta=2$. For the first case the minimum is $\alpha^*=1.27651564...$. We can see the expected features, the derivative tends to zero near $\alpha=\alpha^*$ and $\alpha=2$, and when $\alpha\uparrow2$ the noise intensity decreases sharply towards zero. In the points of parameter space below the curve the system will follow the asymptotic law for $\alpha$-stable noise. Meanwhile, points above the curve will follow Kramer's law. For the second case, since the potential barrier is narrower, the rate of crossing through one jump relative to climbing the barrier is higher, and makes Eq.\ref{eq:weakNoise} have two solutions for any value of $\alpha\in(0,2)$. In this case $\alpha^*$ does not exist and Eq.\ref{eq:minalpha} does not have a solution. In the points of parameter space below the left branch of the curve and to the right of the right branch of the curve the mean escape time will follow the asymptotic law for $\alpha$-stable noise. Meanwhile, in the points between the branches the mean escape time will follow Kramer's law.

The equations presented in this section rely in the hypothesis that Kramer's law describes the diffusion times in the non-asymptotic regime. This is not true for all potentials, for example the asymmetric potential presented in the next section although a exponential law seems to hold.

As a final note, we emphasize that the numerical results do not depend on the potential being Lipschitz or not. When using the Lipschitz potentials we obtained the same curves for the escape time, even though is outside the validity area of the escape time law proven in \cite{imkeller:2006}. This suggests that even if the condition on the derivative of the potential growing superlinearly in $\infty$ is necessary in the mathematical formalism, its shape in a vicinity of the double well is the determinant factor in practice.

\section{HETEROGENEOUS DIFFUSION}

\begin{figure}
    \centering
    \includegraphics[width=12cm]{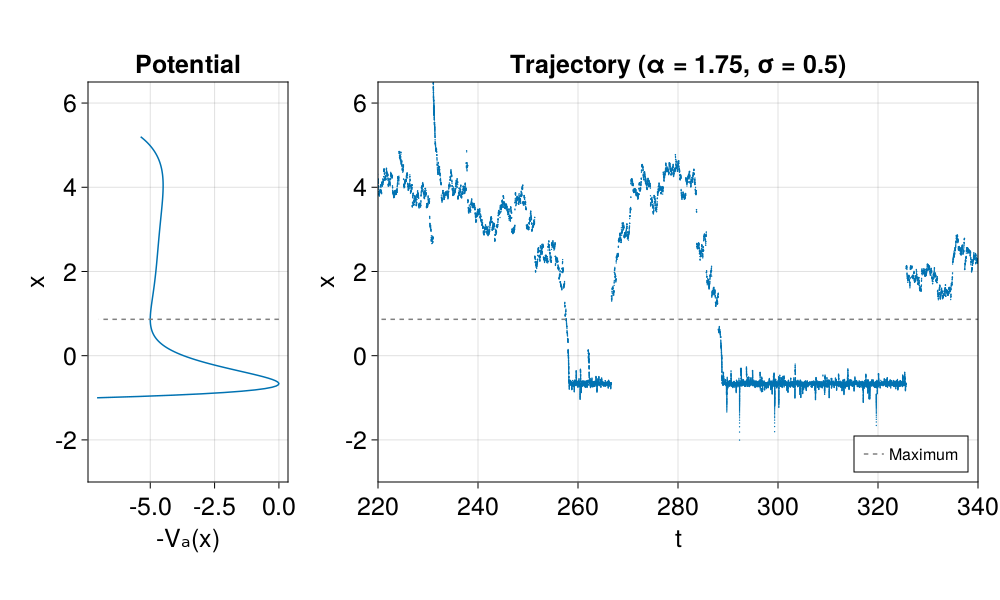}
    \caption{The left panel shows the asymmetric potential (mirrored for clarity) and the right a trajectory. The dashed line marks the local maxima of the potential. The transitions exhibit different behaviours depending on the well they are in, even if the noise parameters do not change. Escape from the left well is realized through one jump, while escapes from the right well can climb the potential barrier in many steps.}
    \label{fig:asymmetricTrajectory}
\end{figure}

From equation \ref{eq:weakNoise} we learn that for a given $\alpha<2$, and a given noise level $\sigma$, whether a system is in the diffusion regime or in the jumping regime depends on the height and width of the barrier from the side of the escape. With a strongly asymmetric potential, this implies that diffusion from one side can be Gaussian-like, determined by an exponential law escape time, while escape from the other side can be due to one large discontinuous jump. We illustrate that in the following. Consider the function 
\begin{align}
    f(x)=2(\tanh(-x/2-1)+1)-x/4 
\end{align}
and the polynomial 
\begin{align}
    g(x)= \frac{5}{2} x^6+\frac{27}{8}x^5-\frac{9}{4}x^4- \frac{45}{8}x^3-3 x^2+5.
\end{align}
The potential defined by $V_a(x)=g(f(x))$ is asymmetric. The left well ($x<0.864$) is deep and narrow ($\omega_L\approx 1.529$ and $\Delta_L= 5$) while the right well ($x>0.864$) is shallow and wide ($\omega_R\approx 3.174$ and $\Delta_R= 0.5$). The process defined with this potential does not follow Kramer's law outside the asymptotic regime even when $\alpha=2$. If the non-asymptotic diffusion takes a similar form to Kramer's time, it should be exponentially high in the left well but comparable to the power law rate in the right well. Figure~\ref{fig:asymmetricTrajectory} shows a trajectory in the asymmetrical potential. While the transitions from the left to the right well are through jumps from the vicinity of the equilibrium, the transitions from right to left can be also be through almost continuous crosses similar to the Gaussian case. 

\begin{figure}
    \centering
    \includegraphics[width=12cm]{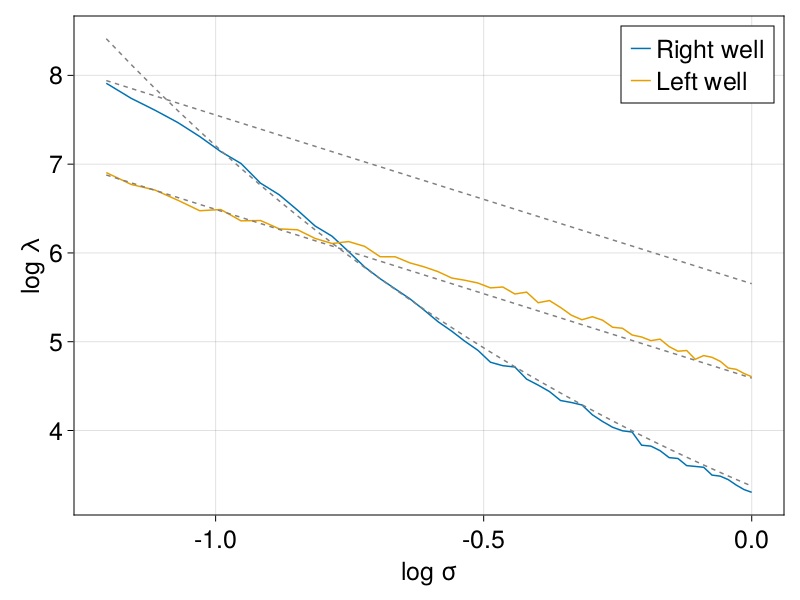}
    \caption{Mean escape times for the two wells of the asymmetrical potential. The two straight dashed lines are the asymptotic escape times for each well, and the curved one is Eq.~\ref{eq:numericalEstimation}. For a wide range of noise intensities, the left well escape times follow a power law but the right well escape times follow approximately the exponential law given by Eq.~\ref{eq:numericalEstimation} before converging to the asymptotic power law.}
    \label{fig:heterogeneousResponse}
\end{figure}

When noise tends to 0 the escape time follows the asymptotic power law given by Eq.\ref{eq:LevyEscapeTimePotential1} in both wells, but with different width. However, over a wide range of noise intensities the response of the process to a change in noise intensity is different in the two wells. Figure~\ref{fig:heterogeneousResponse} shows how for many different values of $\sigma$ the evolution of the average escape time in the left well follows a power law while the escape time in the right well follows an exponential law. For this case, the exponential law is not Kramer's escape rate. A numerical estimation yields 
\begin{align}\label{eq:numericalEstimation}
    \lambda_R \approx \exp{(e^{1.216}\sigma^{-0.759}
    )}.
\end{align}

\section{CONCLUSION}

The purpose of this research is determining how the average escape time switches from strong a dependency on the height of the barrier when $\alpha=2$ (Kramer's law) to a exclusively width dependent law in the case $\alpha<2$, given that the transition of the distributions is continuous. To study this, we first chose a potential such that we can control the height and the width independently. Using scalings, we showed analytically and numerically that there is a simple relationship between the asymptotic average escape times of the whole family of potentials. This allows us to find coordinates based on the parameters such that for all parameters the asymptotic escape time laws share the same graph, a universal curve. 

Then the parameter $\alpha$ was considered. For different values of $\alpha$ we observe that Kramer's law dominates the diffusion rates. This is consistent with the numerical experiments in \cite{chechkin:2005,Chechkin:2007} performed with a potential which is a particular case of our symmetric potential. But this also means that the switching between the height dependency and the width dependency happens for each value of $\alpha$ such that the two rates cross. In particular when $\alpha$ tends to $2$, the two rates always cross, because the prefactor of the power law tends to infinity, causing the rate given by Kramer's law to be below the power law for smaller values of the noise intensity, and the thus the mean escape  time of the $\alpha$-stable process converges pointwise to Kramer's law in a way not described by the asymptotic laws.

This led us to study the level set in which the mean escape time rates are equal, and obtain a formula to compute the minimum value of $\alpha$ such that the crossing exists for a given potential. However, the behaviour of the level set changes with the parameters of the potential, and there are potentials such that both rates intersect always two times, for any value of $0<\alpha<2$.

This has the interpretation that when the probability of crossing the barrier through one jump overtakes the probability of climbing the barrier in several steps, the dependency of the crossing rate on the noise intensity changes as well. This is important for the study of physical systems, for which the noise intensity is not infinitesimally small, and can undergo changes in noise intensity or potential landscape.

There is an apparent contradiction between the universal curve being independent from the parameters of the potential as a geometric object, and the level set that defines the weak noise threshold being dependent, since it also affects the geometry of the curve. The universal curve is proven to work on the asymptotic regimes, while the behaviour of the level set is far from the weak noise limit, since is supposed to approximate where the asymptotic regime breaks down. This makes possible to find potentials such that the weak noise threshold is very different in the two wells, leaving an interval of noise intensities for which the response of the system to a change in the noise intensity follows very different laws. This could affect systems with very asymmetrical potentials, such as in \cite{kirunda2021effects}. Also similar phenomena could appear if the system approaches a saddle-node bifurcation, and the parameters of the well change in a way that the weak noise threshold is crossed, possibly even without changing the noise intensity, giving a change of diffusive mode but in time instead of space. 

In conclusion, motion driven by $\alpha$-stable noise can show two kinds of diffusion. They depend not only on the parameters of the noise process ($\alpha$ and $\sigma$), but also on the potential landscape. The asymptotic laws describe adequately the average escape time when the noise intensity is very small but are of limited use outside this range. In the special case of the symmetric potentials, the two asymptotic laws seem to describe the transition times in both diffusion modes at all noise intensities, but this is not true in general.  

\section{Appendix: $\alpha$-stable processes}
The stochastic process \eqref{eq:Langevin} is an extension of the more familiar Gaussian noise process $dX_t=f(x)dt+\sigma dB_t$. The meaning of the stochastic increment $dB_t$ is that of an infinitely divisible noise process, such that $dB_{t+dt}=\lim_{n\rightarrow \infty}(d\tilde{B}_{t+\delta}+ ....+d\tilde{B}_{t+n\delta})(\sqrt{dt/n})$, with $\delta=dt/n$ and $dB_\tau$ being i.i.d. (independent identically distributed) stochastic variables with zero mean and unit variance. In this case the central limit theorem (CLT) implies that $dB$ is a gaussian random variable with $\langle dB\rangle=0$ and $\langle dB^2\rangle=dt$. We shall not prove the CLT, just remind that a relatively straight forward proof is obtained by use of the characteristic function (CF)
\begin{equation}
cf_X(t)\equiv\langle e^{itX}\rangle=\int_{-\infty}^\infty e^{itx}\phi(x)dx, 
\label{appendix:cf}
\end{equation}
 where $\phi(x)$ is the probability density function (pdf) for $X$.  The characteristic function is thus just the Fourier transform of the pdf. The point of introducing the CF is twofold: Firstly, the distribution for sums of stochastic variables, $Z=X+Y$ involves convolution; $\phi_Z(z)=\int \phi(x)\phi(z-x)dx$, while the Fourier transform of a convolution of two functions is the product of the two Fourier transforms. This is straight forward to generalize for $Z=(X_1+ ... + X_n)/\sqrt{n}$, which gives 
 \begin{equation}
 cf_Z(t)=\prod_{i=1}^n cf_{X_i}(t/\sqrt{n}).
 \label{appendix:sum}
 \end{equation}
Secondly, the characteristic function is the moment generating function for imaginary times:
\begin{equation}
(-i)^n \frac{d}{dt}\langle e^{itX}\rangle |_{t=0}=(-i)^n \int_{-\infty}^\infty \frac{d}{dt}e^{itx}\phi(x)dx|_{t=0}=\langle x^n\rangle.
\end{equation}
With this we may expand the CF; $cf_X(t)=1+i\mu t+i^2\sigma^2t^2/2+{\cal
O}(t^3)$, noticing that the first term is just the integral of the pdf, second term the mean, third term second moment etc. Assuming now $\langle X\rangle=0$, $\langle X^2\rangle=\sigma^2$, higher moments finite and taking $X_i$'s to be i.i.d. we may write \eqref{appendix:sum}
\begin{equation}
    cf_Z(t)=\big(1-\sigma^2 (t/\sqrt{n})^2/2+{\cal O}(t/\sqrt{n})^3\big)^n, 
\end{equation}
and using $\lim_{n\rightarrow \infty}(1+y/n+y_1/n^{3/2}+ ...)^n=e^y$ we get in the limit $n\rightarrow \infty$:

\begin{equation}
cf_Z(t)\rightarrow e^{-\sigma^2t^2/2}, 
\end{equation}
which shows that $Z$ is a gaussian random variable with variance $\sigma^2$, remembering that the gaussian is it's own Fourier transformed (exchanging $\sigma^{-1}$ for $\sigma$).
This is the essence of the CLT. What it tells is that all distributions with finite variance are in the basin of attraction of the gaussian with respect to summing i.i.d. random variables (i.e. taking the mean). A random variable $X$ for which $Z_n=(X_1+...+X_n)$, where all $X_i$'s are drawn from the distribution of $X$, has the same distribution as $c_nX$ is called a stable random variable. Thus, the gaussian random variable is a stable with $c_n=n^{1/2}$. The proof of the CLT relies on the distributions having finite variance, thus if this is not the case the CLT no longer holds. It can be shown that for any stable random variable $X$, there is an $\alpha\in (0, 2]$, such that $Z_n=(X_1+...+X_n)/n^{1/\alpha}$ is a stable distribution \cite{feller}. In analogy to the gaussian case, the stability can be shown from the characteristic function, which in this case can be shown to be $\langle e^{i\theta X}\rangle=e^{-\sigma^\alpha |\theta|^\alpha}$. This is in the case of a symmetric distribution $\phi_\alpha(-x)=\phi_\alpha(x)$, which is what we shall consider here. These are denoted $S\alpha S$ (Symmetric $\alpha$-Stable). Using \eqref{appendix:sum} we have for $S\alpha S$:

\begin{equation}
    cf_Z(\theta)=\langle e^{i\theta(X_1+...+X_n)/n^{1/\alpha}}\rangle=
    (e^{i\theta X/n^{1/\alpha}})^n=(e^{-\sigma^\alpha |\theta|^\alpha /n})^n=e^{-\sigma^\alpha|\theta|^{\alpha}}
\end{equation}

It can be shown that an extension of the CLT can be made stating that sums of any distribution with finite moments $\langle x^\beta\rangle$ of order $0< \beta < \alpha < 2$ and infinite moments of order $\beta \geq \alpha$ will converge to an $\alpha$-stable distribution with stability parameter $\alpha$. It can be shown that $e^{-\sigma^\alpha|\theta|^{\alpha}}$ is indeed a characteristic function, i.e. its inverse Fourier transform is a probability density $\phi_\alpha(x)$ (non-negative, and integrates to one). It can be shown from the explicit form of the CF that the tail of the $\alpha$-stable density distribution behaves as $\phi_\alpha(x)\sim x^{-\alpha-1}$, such that only moments $\langle x^\beta\rangle$, $\beta<\alpha<2$ are finite. The final property needed for a heuristic understanding of \eqref{eq:Langevin} is that $S\alpha S$ stochastic variables can be constructed as the limit of a sum of Poisson jump processes. Thus for $\alpha<2$ the processes are discontinuous. The mathematics necessary for a rigorous treatment is quite involved \cite{samorodnitsky}. Intuitively, the $\alpha$-stable process can be understood as composed of a diffusion process, with randomly distributed jumps. For $0< \alpha\leq 1$, the process is dominated by the jumps, for $1< \alpha < 2$, and as $\alpha\rightarrow 2$ the diffusion process becomes more and more dominant and at $\alpha=2$, the process is continuous.

The fact that the variance is infinite for $\alpha<2$, and even the mean is infinite for $\alpha\leq 1$ can seem strange. Obviously, from any realization of a process, $\{x(\Delta t), x(2\Delta t), ... x(n\Delta t)\}$, the mean $m(n)=(x(\Delta t)+ x(2\Delta t)+ ... +x(n\Delta t))/n$ will be finite. The way the infinity shows up is that $m(n)$ does not converge as $n\rightarrow \infty$. As $n$ grows, jumps so large that it alters the mean will occur. Similarly for the variance.

\subsection*{Acknowledgments}

 We want to thank Alessandro Lovo for many discussions and suggestions and the developers of DrWatson.jl \cite{Datseris2020}, a software package written in Julia used to develop the code with a focus on reproducibility. The code is open source and can be downloaded and used from \cite{GitHub}. This research is supported by the European Union’s Horizon 2020 research and innovation programme under the Marie Skłodowska-Curie Actions grant No.956170 CriticalEarth.

\bibliographystyle{plain}
\bibliography{dblwell}

\end{document}